# Analytical treatment of cold field electron emission from a nanowall emitter

XIZHOU QIN, WEILIANG WANG, NINGSHENG XU, ZHIBING LI*

State Key Laboratory of Optoelectronic Materials and Technologies

School of Physics and Engineering, Sun Yat-Sen University, Guangzhou 510275, P.R. China

RICHARD G. FORBES

Advanced Technology Institute, Faculty of Engineering and Physical Sciences,

University of Surrey, Guildford, Surrey GU2 7XH, United Kingdom

This paper presents an elementary, approximate analytical treatment of cold field electron emission (CFE) from a classical nanowall (i.e., a blade-like conducting structure situated on a flat conducting surface). A simple model is used to bring out some of the basic physics of a class of field emitter where quantum confinement effects exist transverse to the emitting direction. A high-level methodology is presented for developing CFE equations more general than the usual Fowler-Nordheim-type (FN-type) equations, and is applied to the classical nanowall. If the nanowall is sufficiently thin, then significant transverse-energy quantization effects occur, and affect the overall form of theoretical CFE equations; also, the tunnelling barrier shape exhibits "fall-off" in the local field value with distance from the surface. A conformal transformation technique is used to derive an analytical expression for the on-axis tunnelling probability.

These linked effects cause complexity in the emission physics, and cause the emission to conform (approximately) to detailed regime-dependent equations that differ significantly from FN-type equations. In particular, in the emission equations: (a) the zero-field barrier height $H_R$ for the highest occupied state at 0 K is not equal to the local thermodynamic work-function $\phi$, and $H_R$ rather than $\phi$ appears in both the exponent and the pre-exponential; (b) in the exponent, the power dependence on macroscopic field $F_M$ can be $F_M^{-2}$ rather than $F_M^{-1}$ (or can be something intermediate); and (c) in the pre-exponential the explicit power dependences on $F_M$ and $H_R$ are different from those associated with the elementary FN-type equation, and a different universal constant appears. Departures of this general kind from FN-like behaviour should be expected whenever quantum-confinement effects influence field electron emission. FN-type equations can be recognized as the CFE equations that apply when no significant nanoscale quantum confinement effects occur.

*Author for correspondence (stslzb@mail.sysu.edu.cn)

## 1. Introduction

This paper derives theoretical equations (for line and area emission current density) for cold field electron emission (CFE) from a "classical nanowall". The "metallic" nanowall studied is an elementary model for a larger group of quasi-two-dimensional materials, including those built from a few layers of graphite. The aim here is to understand the basic physics of CFE from nanowalls A longer-term aim is to develop more sophisticated nanowall equations, for instance by incorporating atomic-structure effects.

For clarity, we state some definitions. "Fowler-Nordheim (FN) tunnelling" (Fowler & Nordheim 1928) is field-induced electron tunnelling through a potential energy (PE) barrier that is exactly or approximately triangular. "Cold field electron emission (CFE)" is the statistical electron emission regime in which: (a) the electrons inside and close to the emitting surface of a condensed-matter field emitter are (very nearly) in thermodynamic equilibrium; and (b) most emitted electrons escape by FN tunnelling from states close to the emitter Fermi level. "Fowler-Nordheim-type (FN-type) equations" are a family of approximate equations that describe CFE from "bulk metals".

For CFE theory, the essential characteristic of a "bulk metal" is that the electron states in the emitter's conduction band may be treated as travelling-wave states for all directions inside the emitter. This implies that all dimensions of the emitter must be large enough for there to be no significant standing-wave effects (quantum confinement effects) in any direction, though there will, of course, be interference effects associated with back reflection from the emitting surface.

A "nanowall" is a blade-like material structure that stands upright on a substrate. It has a width that is small absolutely and small in comparison with its height, and has a length that is large or very large in comparison with its width. Carbon nanowalls can be grown in suitable deposition conditions, and can be efficient field electron emitters (Wu et al. 2002, Teii & Nakashima 2010). Present nanowall emitters tend to grow as irregular honeycombe-type structures containing many short interconnected nanowalls. But, if a longer, more uniform, single nanowall could be grown, this structure might have interesting applications as a line electron source. Thus, it is felt that a theory of CFE from nanowalls may have both scientific and technological relevance.

For CFE theory, the essential feature of a nanowall is that its width is sufficiently small that the solutions of the Schrödinger equation in the direction of the width are standing waves. In consequence, the electron energy component for this direction is quantized. As shown below, CFE from a nanowall does not obey the usual forms of FN-type equation (except, perhaps, in a very generalised sense). It seems better to regard the equations describing CFE from nanowalls as members of a different (but related) family of theoretical CFE equations. Our aim is to derive elementary equations belonging to the family of "nanowall" CFE equations.

To help focus on the effects (quantization and barrier-field fall-off) related to small nanowall width, and to make the problem analytically tractable, simplifications are made. As shown in Fig. 1, we assume the nanowall has a simple profile, and is uniform in the "long" direction. We disregard its atomic structure, assume that its surface has uniform local work-function, and disregard the possibility of field penetration and band-bending. This means that the nanowall can be treated as a classical conductor with the shape shown in Fig. 1: we call this a "classical nanowall". Inside the nanowall, we use a free-electron model. We also disregard the correlation-and-exchange ("image-type") interaction between departing electrons and the nanowall: this means the potential-energy (PE) barrier at the nanowall surface can be derived by classical electrostatic arguments. FN made equivalent simplifications in their treatment of CFE from a bulk metal with a planar surface.

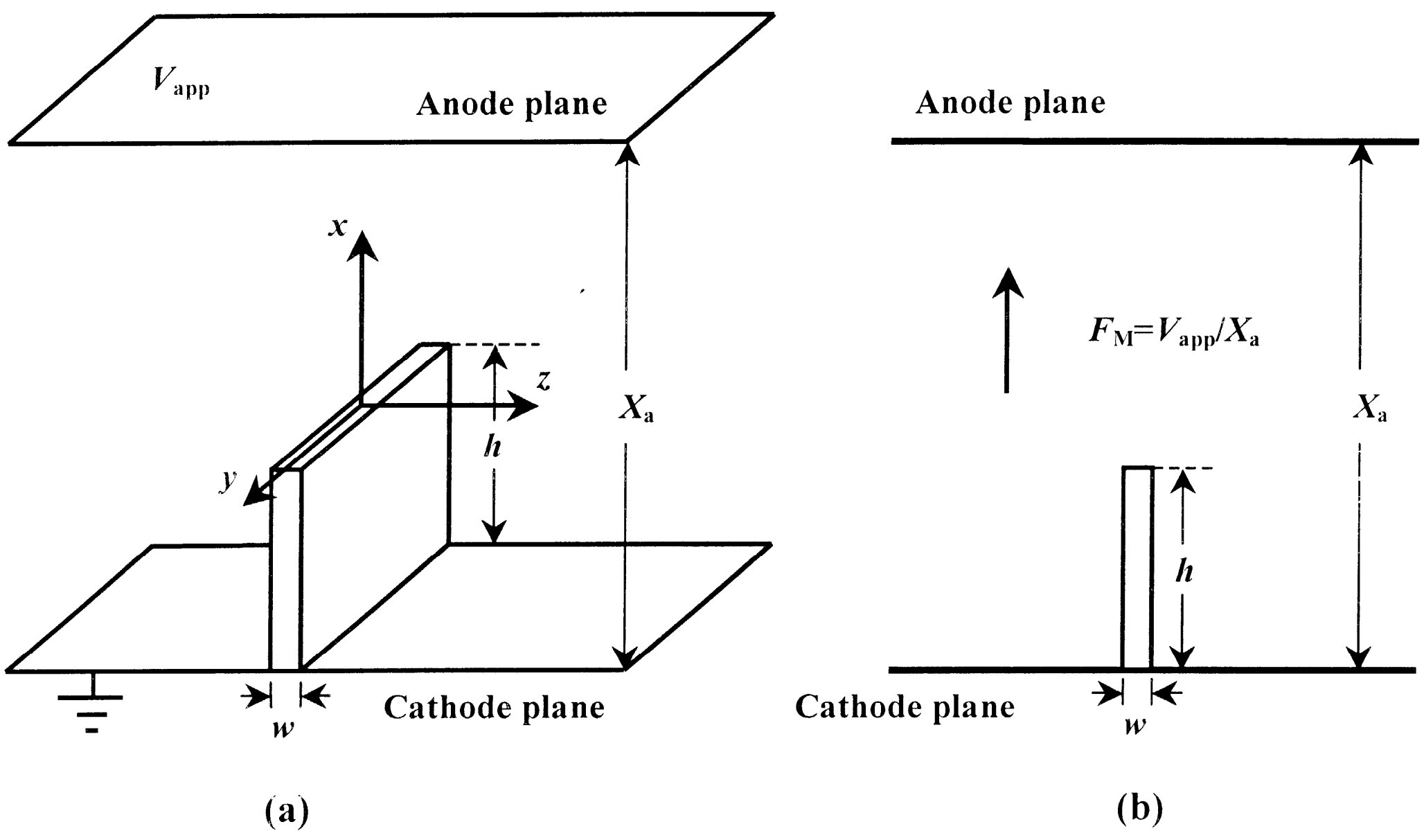

Fig. 1. Classical nanowall: (a) three-dimensional view; (b) projection onto the $z$-$x$ plane.

FN solved the Schrödinger equation exactly for an exact triangular PE barrier. However, in our model, the barrier above the nanowall top surface is not exactly triangular. When barriers are approximately but not exactly triangular, it is usually mathematically impossible to solve the Schrödinger equation exactly in terms of the common functions of mathematical physics. Therefore, we use the simple-JWKB formula (Jeffreys 1925, also see Forbes 2008b) to generate an approximate solution. This approach has been widely used in CFE theory.

When (as here) the nanowall is standing on one of a pair of parallel plates, between which an uniform "macroscopic field" $F_M$ would exist in its absence, a "field enhancement factor" $\gamma$ can be defined by

$$F = \gamma F_{\mathrm{M}} = \gamma V_{\mathrm{app}} / X_{\mathrm{a}} , \qquad (1.1)$$

where $X_a$ is the plate separation, $V_{app}$ is the voltage between them, and $F$ is a local electric field at the emitter surface, called the "barrier field".

These assumptions mean that the equivalent FN-type equation (for comparison with results here) is the elementary FN-type equation, expressed in terms of $F_M$ (see Forbes 2004). This equation gives the emission current density $J_A$ (emission current per unit area) at zero temperature, as a function of $F_M$ and the local thermodynamic work-function $\phi$, and can written in the equivalent forms:

$$J_{\mathrm{A}} = a\phi^{-1}\gamma^2 F_{\mathrm{M}}^2 \exp[-b\phi^{3/2} / \gamma F_{\mathrm{M}}] = z_{\mathrm{S}} (d_{\mathrm{F}}^{\mathrm{el}})^2 \exp[-b\phi^{3/2} / \gamma F_{\mathrm{M}}] . \qquad (1.2)$$

Here, $a$ and $b$ are the first and second FN constants, and $z_S$ is the Sommerfeld supply density, as defined in Table I. $d_F^{el}$ is (the elementary approximation for) a parameter called the "decay width at the Fermi level", discussed below. The second form is less commonly used than the first, but is a more fundamental form that arises naturally during the derivation of (1.2) (see Forbes 2004).

Table 1. Basic universal emission constants. Values are given in the units often used in field emission.

| Name | Symbol | Derivation | Expression | Numerical value | Units |
|---|---|---|---|---|---|
| Sommerfeld supply density[a] | $z_S$ | - | $4\pi e m_e / h_P^3$ | $1.618311 \times 10^{14}$ | A m$^{-2}$ eV$^{-2}$ |
| JWKB constant for electron | $g_e$ | - | $4\pi(2m_e)^{1/2}/h_P$ | 10.24624 | eV$^{-1/2}$ nm$^{-1}$ |
| First Fowler-Nordheim constant | $a$ | $z_S(e/g_e)^2$ | $e^3/8\pi h_P$ | $1.541434 \times 10^{-6}$ | A eV V$^{-2}$ |
| Second Fowler-Nordheim constant | $b$ | $2g_e/3e$ | $(8\pi/3)(2m_e)^{1/2}/eh_P$ | 6.830890 | eV$^{-3/2}$ V nm$^{-1}$ |

[a]The supply density is the electron current crossing a mathematical plane inside the emitter, per unit area of the plane, per unit area of energy space, when the relevant electron states are fully occupied. In a free-electron model for a bulk conductor, the supply density is isotropic, is the same at all points in energy space, and is given by $z_S$.

The paper's structure is as follows. Section 2 outlines a general method deriving theoretical CFE

equations. Section 3 then describes our nanowall model, and discusses issues relating to electron supply. Section 4 derives a formal expression for nanowall current density, and compares this with the elementary FN-type equation. Section 5 derives the electrostatic potential variation in the symmetry plane, by means of a conformal transformation. Section 6 derives detailed theoretical equations for CFE from a nanowall. Section 7 provides a summary and discussion. Some mathematical details are relegated to Appendix A.

In this paper, $e$ denotes the elementary positive charge, $m_e$ the mass of an electron, $h_P$ Planck's constant and $k_B$ Boltzmann's constant. We also use the universal emission constants defined in Table 1, and the usual electron emission convention that fields, currents and current densities are treated as positive, even though they would be negative in classical electrostatics. Note, however, that the quantity $V(x)$ below is conventional electrostatic potential, so the corresponding field is defined by $\mathrm{d}V/\mathrm{d}x$ rather than by $-\mathrm{d}V/\mathrm{d}x$. The theory is constructed within the framework of the International System of Quantities (ISQ) (BIPM 2006), but—where appropriate—uses (instead of SI base-units) the customary units of field emission. (These customary units simplify formula evaluation.)

## 2. The structure of theoretical CFE equations

This section develops further a methodology used by one of us (Forbes 2004), so that it can describe the common general structure of equations that apply to cold field electron emission. In this section the barrier field $F$ is used as the independent field-like variable.

To establish an expression for local emission current density (ECD) (either a current per unit area

$J_A$, or a current per unit length $J_L$), three basic steps are needed. (1) A reference electron state "R" in the emitter is chosen, and the zero-field height $H_R$ (see below) of the barrier seen by an electron in this state is identified. (2) The escape probability (tunnelling probability) $D_R$ and decay width $d_R$ for the barrier seen by an electron in state R are calculated, usually by an approximate quantum-mechanical method. (3) The ECD contributions of all relevant electron states (particularly those energetically close to R) are calculated, and the result of summing the contributions is written in the form

$$J(F,T) = Z_R(F,T) D_R(F) \tag{2.1}$$

where $T$ is the emitter's thermodynamic temperature, $J(F,T)$ is the ECD, and $Z_R(F,T)$ is called the "effective supply of electrons for state R".

Details of calculating $Z_R$ are different for different families of theoretical CFE equations, but theory concerning the tunnelling probability $D_R$ is similar for all families. When emission comes from a set of "electronic sub-bands", an analogous procedure is applied to each sub-band, and the total current density is obtained by summing contributions from the various sub-bands. (But, in some cases, emission from one sub-band may dominate.)

### *(a) Tunnelling probability*

A tunnelling barrier is described by an energy function $M(l,F,H)$, where $l$ is distance measured from some convenient reference point, $F$ is the barrier field, and $H$ (the "zero-field barrier height") is the barrier height seen by an approaching electron when $F$=0. Physically, $M(l,F,H)$ is given by $M$=$U$–$E_{\mathrm{n}}$, where $U$ is the PE that appears in the Schrödinger equation, and $E_{\mathrm{n}}$ is the total electron-energy component associated with motion in the direction of tunnelling, referred to the same energy-zero as $U$ ($E_{\mathrm{n}}$ is called the "forwards energy", here). The barrier is the region where $M$≥0.

A parameter $G(F,H)$, called here the "JWKB exponent", is then defined by

$$G(F,H) \equiv g_{\mathrm{e}} \int \sqrt{M(l,F,H)}\,\mathrm{d}l\,, \tag{2.2}$$

where $g_{\mathrm{e}}$ is the so-called "JWKB constant for an electron", defined in Table 1, and the integral is taken over the region where $M$≥0. Both $M$ and $G$, and other functions below, may be functions of additional variables that describe the geometry of the emitter, but these dependences are not shown.

The "exact triangular (ET) barrier" used in the derivation of (1.2) is

$$M^{\mathrm{ET}}(l,F,H) = H - eFl\,, \tag{2.3}$$

where $l$ here is measured from the emitter's "electrical surface" (Lang & Kohn 1973, Forbes 1999), which is "the surface where the field starts". In the case of a classical conductor, the electrical surface is the conductor surface. For the ET barrier, the JWKB integral (2.2) is taken over the region $0 \leq l \leq H/eF$, and the well-known result is

$$G^{\mathrm{ET}}(F,H) = bH^{3/2} / F\,. \tag{2.4}$$

For barriers that are only approximately triangular, the JWKB exponent is written

$$G = \nu G^{\mathrm{ET}} = \nu b H^{3/2} / F\ , \tag{2.5}$$

where $\nu$ ("nu") is a factor that corrects the tunnelling exponent because the barrier shape is not exactly triangular. $\nu$ is derived by evaluating integral (2.2).

For the CFE regime, a suitable formal expression for the tunnelling probability $D(F,H)$ is

$$D(F,H) \approx P\exp[-G] = P\exp[-\nu b H^{3/2} / F] \tag{2.6}$$

where $P$ is a "tunnelling prefactor" discussed in Forbes 2008b. Except in a few special cases, the calculation of $P$ requires specialist mathematical/numerical techniques. These are only just beginning to be applied to the one-dimensional barriers commonly discussed in practical CFE applications; thus, a very common approximation (also used here) has been to put $P\rightarrow 1$.

In CFE theory it is important to know how quickly the tunnelling probability varies as the zero-field height $H$ of the barrier changes. This variation can be described by either a "decay rate" $k_{\mathrm{d}}$ or a "decay width" $d$ ($\equiv 1/ k_{\mathrm{d}}$) defined by:

$$1/d \equiv k_{\mathrm{d}} \equiv -\partial \ln D / \partial H = -\partial \ln P / \partial H + \partial G / \partial H\ . \tag{2.7}$$

The literature uses both decay rate (e.g., the Modinos (1984) $c_0$, or the Jensen (2007) $\beta_{\mathrm{F}}$) and decay width (e.g., the Gadzuk and Plummer (1973) $d$). Decay width $d$ (normally measured in eV) is preferred here, because it has the same dimensions as the Boltzmann factor $k_{\mathrm{B}}T$, and the dimensionless ratio ($k_{\mathrm{B}}T/d$) is important in CFE theory. The physical interpretation is that the tunnelling probability $D$ decreases by a factor of approximately e (≅2.7) when $H$ increases by $d$.

When deriving the elementary FN-type equation (1.2), we put $P\to 1$, $\nu\to 1$, $G\to bH^{3/2}/F$, and obtain the "elementary" variants of $d$ and $k_d$ as

$$1/d^{\mathrm{el}} = k_{\mathrm{d}}^{\mathrm{el}} = \tfrac{3}{2} bH^{1/2}/F = g_{\mathrm{e}} H^{1/2}/eF\,, \tag{2.8}$$

where the derivation of $b$ has been used (see Table 1). In more general situations, the outcome of definition (2.7) is written

$$d^{-1} = k_{\mathrm{d}} \equiv \tau k_{\mathrm{d}}^{\mathrm{el}} = \tau g_{\mathrm{e}} H^{1/2}/eF\,, \tag{2.9}$$

where $\tau$ is a "decay-rate correction factor" defined by (2.9), and obtained by evaluating (2.7). The factor $\tau$ as used here can be understood as a generalization both of the factor $t$ that appears in the Standard FN-type equation (Murphy & Good 1956), and of the factor $\tau$ that appears in partially generalized CFE theories where $P$ is not taken into account (e.g., Forbes 2004).

When deriving theoretical CFE equations, one needs the values of tunnelling probability and decay width for the barrier related to the reference state of interest. Labelling quantities specific to this state/barrier by the subscript "R", and also substituting $F=\gamma F_{\mathrm{M}}$ in (2.4) and (2.9), we get

$$D_{\mathrm{R}}(F) \approx P_{\mathrm{R}} \exp[-G_{\mathrm{R}}] = P_{\mathrm{R}} \exp[-\nu_{\mathrm{R}} bH_{\mathrm{R}}^{3/2}/\gamma F_{\mathrm{M}}]\,, \tag{2.10}$$

$$d_{\mathrm{R}} \approx (\tau_{\mathrm{R}}^{-1}\gamma)(e/g_{\mathrm{e}}) H_{\mathrm{R}}^{-1/2} F_{\mathrm{M}}\,, \tag{2.11}$$

For states with zero-field barrier height $H$ close to $H_R$, a Taylor-type expansion allows the tunnelling probability $D(F,H)$ to be approximated. If only the linear term is used (which is the customary approximation), then this is given by

$$D(F,H) = D_R(F)\exp[-(H - H_R)/d_R]. \tag{2.12}$$

*(b) Electron supply*

The emitter's electron states can be labelled by a set of quantum numbers. In the free-electron-type models often used, there will be three directional quantum numbers and a spin quantum number. The symbol $\boldsymbol{Q}$ labels an arbitrary individual member of this four-component set. Each internal electron state $\boldsymbol{Q}$ that approaches the emitter surface from the inside makes a contribution $j_{\boldsymbol{Q}}$ to the total ECD $J$, with $j_{\boldsymbol{Q}}$ given by

$$j_{\boldsymbol{Q}} = e\Pi_{\boldsymbol{Q}} f(E_{\boldsymbol{Q}},T)D(F,H_{\boldsymbol{Q}}). \tag{2.13}$$

Here $e\Pi_{\boldsymbol{Q}}$ is the electron-current-density component (associated with state $\boldsymbol{Q}$) that approaches the emitting surface in a direction normal to the surface, and $f(E_{\boldsymbol{Q}},T)$ is the occupancy of state $\boldsymbol{Q}$.

A formula for the total ECD $J$ is obtained by summing the contributions from all relevant individual states:

$$J = \sum_{\boldsymbol{Q}} j_{\boldsymbol{Q}} = e \sum_{\boldsymbol{Q}} \Pi_{\boldsymbol{Q}} f(E_{\boldsymbol{Q}}, T) D(F, H_{\boldsymbol{Q}}) . \tag{2.14}$$

Making use of the Taylor expansion (2.12), this can be put in the form of (2.1), with

$$Z_{\mathrm{R}}(F,T) = e \sum_{\boldsymbol{Q}} \Pi_{\boldsymbol{Q}} f(E_{\boldsymbol{Q}}, T) \exp[-(H_{\boldsymbol{Q}} - H_{\mathrm{R}}) / d_{\mathrm{R}}] . \tag{2.15}$$

The summation then has to be evaluated for the system under investigation. The form of the result will depend on the emitter geometry, since (for emitters that are "small on the nanoscale" in any direction) this will affect the way in which the emitter's internal states need to be quantized, even with free-electron models.

For the bulk metal case, in thermodynamic equilibrium at zero temperature, the results are particularly simple (as long known). For the relevant reference state "F" (sometimes called the "forwards state at the Fermi level") an electron at the Fermi level approaches the emitting surface normally. The zero-field barrier height is equal to the local thermodynamic work-function $\phi$. The summation in (2.15) can be carried out in several alternative ways. In particular, it can be reduced to a double integral in a "P-T energy space" (see Forbes 2004) in which the variables of integration are the electron's total energy and its kinetic energy parallel to the emitter surface. The two integrations introduce a factor $d_{\mathrm{F}}^2$ into the result. Forbes (2004) shows that

$$Z_{\mathrm{R}}(F, T=0) \rightarrow Z_{\mathrm{F}}(F) \approx z_{\mathrm{S}} d_{\mathrm{F}}^2 , \tag{2.16}$$

where $z_{\mathrm{S}}$ is the Sommerfeld supply density as defined in Table 1, and $d_{\mathrm{F}}$ is the decay width for reference state F (often called the "decay width at the Fermi level"). In the elementary FN-type equation, $Z_F$ is obtained by replacing $d_F$ by $d_F^{el}$.

In the derivation of FN-type equations, the parameters associated with an exact triangular barrier and the elementary FN-type equation play a special role. This is because FN tunnelling is defined as tunnelling through an approximately triangular barrier. Mathematically, it is convenient to use correction factors to relate more sophisticated FN-type equations to the elementary equation. It is also convenient to treat other families of CFE equations by introducing correction factors.

## 3. Electron supply for a classical nanowall

For quantum-mechanical purposes, the classical nanowall shown in Fig. 1 can be modelled (in respect of the $z$-direction) as a deep rectangular PE well. In this paper, the total energy and energy components are measured relative to the energy level of the well base, and are denoted by the basic symbol $W$. Thus, an electron confined in the classical nanowall has total energy $W$ given by

$$W = W_x + W_y + W_z \,, \tag{3.1}$$

where $W_i$ ($i$= $x,y,z$) is the energy component for motion along the $i$-axis. For the $x$ and $y$ directions the motion is free, with

$$W_x = p_x^2 / 2m_e, \quad W_y = p_y^2 / 2m_e \,, \tag{3.2}$$

where $p_x$ and $p_y$ are the momenta in the $x$ and $y$ directions, respectively.

The energy component $W_z$ is quantized, and its allowed values are given approximately by the values for an infinitely deep PE well, namely

$$W_z \approx \frac{1}{2m_\mathrm{e}}\left(\frac{nh_\mathrm{P}}{2w}\right)^2 \equiv W_{zn} = n^2 W_{z1}\,, \tag{3.3}$$

where $w$ is the well width, $n$ is a positive integer that defines the electron's "vibrational level", and $W_{zn}$ and $W_{z1}$ are defined by this equation ($W_{z1}$ is the value of $W_z$ associated with the $n$=1 level). This quantization of $W_z$ means that the electron states in the nanowall are split into sub-bands SB1, SB2 … SB$n$ . . ., etc., with each sub-band associated with a particular value of $n$.

Since (in the CFE regime) the emitter is treated as very nearly in thermodynamic equilibrium, the occupancy of electron state $\boldsymbol{Q}$ is given by the Fermi-Dirac distribution function

$$f(W_\mathbf{Q},T) = \frac{1}{1+\exp[(W_\mathbf{Q} - W_\mathrm{F})/k_\mathrm{B}T]} \tag{3.4}$$

Here, $W_\mathrm{F}$ is the Fermi energy (i.e., the total energy of an electron at the Fermi level). For electrons in sub-band SB$n$, the Fermi-Dirac occupation probability is given by the mathematical expression

$$f_n = \frac{1}{1+\exp[\{W_x + (p_y^2/2m_\mathrm{e}) + W_{zn} - W_\mathrm{F}\}/k_\mathrm{B}T]}\,. \tag{3.5}$$

For sub-band SB$n$, the number of electrons from SB$n$ crossing a plane normal to the $x$-axis per unit time, per unit length of the nanowall (in the $z$-direction), per unit range of the energy-component $W_x$, is called the "line supply function for sub-band SB$n$" and is denoted by $N_{L,n}(W_x,T)$. This function is obtained by integrating over the allowed range of values of $p_y$, taking into account the occupancy of the electron states concerned. Thus:

$$N_{\mathrm{L},n}(W_x,T) = (2/h_\mathrm{P}^2)\int_{-\infty}^{+\infty} f_n dp_y \,. \tag{3.6}$$

$N_{L,n}(W_x,T)$ has the SI units (s$^{-1}$ m$^{-1}$ eV$^{-1}$), and is not the same physical quantity as the (area) supply function (Nordheim 1928) that appears when FN-type equations are derived by integration via the normal energy distribution.

## 4. Formal expressions for nanowall current density

### *(a) Tunnelling barrier parameters*

In the determination of $G$, the JWKB integral (2.2) has to be evaluated along an appropriate path, which may in principle be curved and will be influenced by the pattern of electric field lines in the region through which the electron passes. To avoid the complications associated with curved paths of integration (Kapur & Peierls 1937), but still provide a result illustrative of nanowall emission, we consider a straight-line path of integration that starts from a symmetry position S on the nanowall top surface and coincides with the $x$-axis. The zero-point of the $x$-axis is taken at the nanwall top surface, at S, and the barrier field is defined as the local surface field $F_S$ at position S. This field $F_S$ is related to the macroscopic field $F_M$ by (1.1), which defines a corresponding field enhancement factor $\gamma_S$ that is independent of $F_M$ and is discussed further below. In what follows, it is convenient to show field dependences as a function of $F_M$ rather than $F_S$.

Fig. 2 shows schematically the variation, along the $x$-axis, of the electron PE $U(x,F_M)$ (measured relative to the base of the PE well). In the present treatment (which disregards image-type effects), $U(x,F_M)$ is given by

$$U(x,F_M) = W_o - eV(x,F_M) = W_F + \phi - eV(x,F_M) \tag{4.1}$$

where $W_o$ is the nanowall inner PE (or "electron affinity"), $\phi$ is the local thermodynamic work-function of its surface, and $V(x,F_M)$ is the conventional electrostatic potential in the surrounding space, taking the nanowall potential as zero. Obviously, the "shape" of $V(x,F_M)$ depends on the nanowall shape, and the size of $V(x,F_M)$ depends on the value of $F_M$.

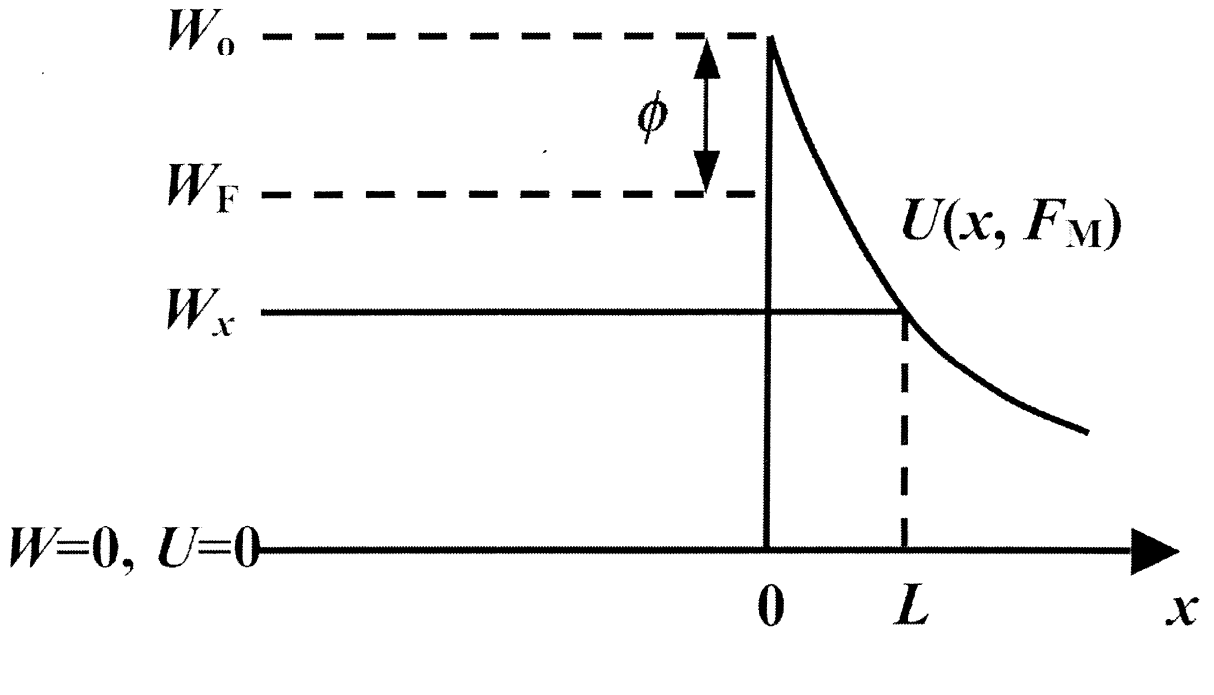

Fig. 2. Schematic diagram showing how the electron potential energy varies along the $x$-axis.

For states with forwards energy $W_x<W_o$, there exists a tunnelling barrier of zero-field height $H$ given by

$$H = W_o - W_x ,\qquad (4.2)$$

Thus, any formula above involving $H$ can alternatively be written in terms of $W_x$. Due to the disregard of image-type effects, the barrier's inner edge is at $x$=0. Its outer edge is at $x$=$L$, where

$$eV(L, F_M) = H = W_o - W_x .\qquad (4.3)$$

In the CFE regime, (2.6) is an adequate expression for tunnelling probability. With the further approximation $P \rightarrow 1$, the tunnelling probability $D(W_x,F_M)$ for an electron approaching the nanowall top surface with forwards energy $W_x$ becomes

$$D(W_x, F_M) \approx \exp[-G(W_x, F_M)] .\qquad (4.4)$$

$D(W_x,F_M)$ decreases exponentially as $W_x$ decreases; thus, most of the emission comes from occupied states with $W_x$ as high as possible. However, in the CFE regime the occupancy of states falls off sharply above the Fermi level. Thus, most of the emission comes from states near the Fermi level. For states at the Fermi level, the largest possible value of $W_x$ is

$$W_{xR1} = W_F - W_{z1} = W_o - \phi - W_{z1} ,\qquad (4.5)$$

which corresponds to a state with $n$=1 and $p_y$=0. This can be specified as the reference state ("R1") for sub-band SB1; $W_{x\mathrm{R}1}$ is the corresponding forwards energy.

More generally, for states at the Fermi level, the largest possible value of forwards energy $W_x$ for states in sub-band SB$n$ is the forwards energy $W_{x\mathrm{R}n}$ for the corresponding reference state R$n$:

$$W_{x\mathrm{R}n} \;=\; W_{\mathrm{F}} - W_{zn} \,. \tag{4.6}$$

The JWKB exponent for the barrier related to state R$n$ is denoted by $G_n(F_{\mathrm{M}})$ and given by

$$G_n(F_{\mathrm{M}}) = g_{\mathrm{e}} \int_0^{x_n} \sqrt{(\phi + W_{zn}) - eV(x,F_{\mathrm{M}})}\,\mathrm{d}x \,. \tag{4.7}$$

$x_{\mathrm{n}}$ is the outer classical turning point for the barrier seen by an electron in state R$n$, and is found by solving (4.3), with $L$=$x_n$, $H$=$\phi$+$W_{zn}$.

Thus, in respect of its escape by tunnelling, an electron in state R$n$ sees as barrier of zero-field height $H_n$ given by

$$H_n = \phi + W_{zn} \approx \phi + n^2 W_{z1} \,. \tag{4.8}$$

The reason that $H_n$ is greater than the local thermodynamic work-function $\phi$ is that the electron has to "use up" some of its kinetic energy to provide the energy $W_{zn}$ associated with its localization in the $z$-direction. Obviously, if $W_{zn}$>$W_{\mathrm{F}}$, then the sub-band is empty (except for any electrons that may get into it as a result of thermal activation), and contributes little or no emission.

Since we have put $P\to 1$ in our model, the decay width $d_n$ for reference state R$n$ is given [using eq. (9) and noting d$W_x$= –d$H$] by the approximate expression

$$d_n^{-1} \approx -\left.\frac{\partial G}{\partial W_x}\right|_{W_x=W_{x\mathrm{R}n}} = (g_\mathrm{e}/2)\int_0^{x_n} \frac{1}{\sqrt{\phi + W_n - eV(x,F_\mathrm{M})}}\mathrm{d}x\,. \tag{4.9}$$

*(b) Characteristic line current density*

The present model derives an expression for a quantity $J_\mathrm{L}(F_\mathrm{M},T)$ called the "characteristic line current density". $J_\mathrm{L}(F_\mathrm{M},T)$ is the electron current that would be emitted per unit length of the nanowall if each electron arriving at its top surface had an escape probability corresponding to the barrier defined for symmetry position S. For the present paper this is an adequate approximation (more sophisticated treatments would introduce a correction factor). The contribution $J_{\mathrm{L},n}(F_\mathrm{M},T)$ from sub-band SB$n$ to $J_\mathrm{L}(F_\mathrm{M},T)$ is

$$J_{\mathrm{L},n}(F_\mathrm{M},T) = e\int_{-\infty}^{+\infty} D(W_x,F_\mathrm{M})N_{\mathrm{L},n}(W_x,T)\mathrm{d}W_x\,, \tag{4.10}$$

where the lower limit on $W_x$ can be taken as $-\infty$ because $D(W_x,F_\mathrm{M})$ gets exponentially smaller as $W_x$ decreases.

For reference state R$n$, the difference $-(H-H_\mathrm{R})$ in (2.12) can alternatively be written

$$-(H-H_n) = W_x - W_{x\mathrm{R}n} = W_x + W_{zn} - W_\mathrm{F}\,. \tag{4.11}$$

Combining (2.12), (3.5), (3.6), (4.1) and (4.11) yields

$$J_{\mathrm{L},n}(F_{\mathrm{M}},T)=\frac{2e}{h_{\mathrm{P}}^{2}}\exp[-G_{n}]\int_{-\infty}^{\infty}\int_{-\infty}^{\infty}\frac{\exp\left[d_{n}^{-1}(W_{x}+W_{zn}-W_{\mathrm{F}})\right]}{1+\exp[(k_{\mathrm{B}}T)^{-1}(W_{x}+p_{y}^{2}/2m_{\mathrm{e}}+W_{zn}-W_{\mathrm{F}})]}\mathrm{d}p_{y}\mathrm{d}W_{x}, \qquad (4.12)$$

By reversing the order of integration, and defining $u=W_{x}+p_{y}^{2}/2m_{\mathrm{e}}+W_{zn}-W_{\mathrm{F}}$, the double integral reduces to

$$I=\int_{-\infty}^{\infty}\exp[-p_{y}^{2}/2m_{\mathrm{e}}d_{n}]\int_{-\infty}^{\infty}\frac{\exp[d_{n}^{-1}u]}{1+\exp[(k_{\mathrm{B}}T)^{-1}u]}\mathrm{d}u\,\mathrm{d}p_{y}\ . \qquad (4.13)$$

The integral in $u$ is a standard form that results in a temperature-dependent correction factor $(\pi k_{\mathrm{B}}T)/\sin(\pi k_{\mathrm{B}}T/d_{n})$, and the resulting integral over $p_y$ yields the extra factor $(2\pi m_{\mathrm{e}}d_{n})^{1/2}$. Thus, (4.13) reduces to

$$J_{\mathrm{L},n}(F_{\mathrm{M}},T)=z^{\mathrm{nw}}\{(\pi k_{\mathrm{B}}T/d_{n})/\sin(\pi k_{\mathrm{B}}T/d_{n})\}\ d_{n}^{3/2}\ \exp[-G_{n}], \qquad (4.14)$$

where $z^{\mathrm{nw}}$ is an universal constant with the value

$$z^{\mathrm{nw}}\equiv(2e/h_{\mathrm{P}}^{2})(2\pi m_{\mathrm{e}})^{1/2}\cong1.119769\times10^{5}\ \ \mathrm{A\,m^{-1}\,eV^{-3/2}} \qquad (4.15)$$

The proof of (4.14) is valid within the CFE emission regime (though not at higher temperatures), and in this regime the temperature-dependent correction factor is always relatively small. The error in omitting it is always far less than the error that has been involved in disregarding image-type effects.

Thus, practical calculations can use the zero-temperature limiting form

$$J_{\mathrm{L},n}(F_{\mathrm{M}}) \approx z^{\mathrm{nw}} d_n^{3/2} \exp[-G_n]. \tag{4.16}$$

The characteristic line current density $J_{\mathrm{L}}(F_{\mathrm{M}})$ is obtained by summing the contributions from the various sub-bands. If the width of the nanowall is sufficiently small, then a working field range will exist where $J_{\mathrm{L},n} << J_{\mathrm{L},1}$ for all $n \geq 2$, and emission from the $n=1$ sub-band will be dominant. We call this the "thin-nanowall case". The requirement $(G_2 - G_1) > 1$ is a suitable condition for this to occur, and is discussed further in Section 7. In such circumstances, we can neglect emission from the higher sub-bands and take $J_{\mathrm{L}}(F_{\mathrm{M}}) \approx J_{\mathrm{L},1}(F_{\mathrm{M}})$. In what follows, it is assumed that the nanowall is sufficiently thin to allow this.

At the other limit, as the nanowall width gets larger, the sub-bands with $n>1$ contribute increasingly to the emission, and the theory ultimately becomes equivalent to that associated with the elementary FN-type equation.

*(c) Formal comparison with elementary FN-type equation*

In the thin nanowall case, we can derive a "characteristic area current density" $J_{\mathrm{A}}^{\mathrm{tnw}}(F_{\mathrm{M}})$ by dividing $J_{\mathrm{L}}(F_{\mathrm{M}})$ by the nanowall width $w$, and then using (2.10) and (2.11) to expand the resulting expression:

$$J_{\mathrm{A}}^{\mathrm{tnw}}(F_{\mathrm{M}}) \approx z^{\mathrm{nw}} w^{-1} d_1^{3/2} \exp[-G_1] \tag{4.17a}$$

$$= z^{\mathrm{nw}} w^{-1} d_1^{3/2} \exp[-\nu_1 b H_1 / \gamma_{\mathrm{S}} F_{\mathrm{M}}] \tag{4.17b}$$

$$= a^{\mathrm{nw}} w^{-1} \tau_1^{-3/2} H_1^{-3/4} \gamma_{\mathrm{S}}^{3/2} F_{\mathrm{M}}^{3/2} \exp[-\nu_1 b H_1 / \gamma_{\mathrm{S}} F_{\mathrm{M}}] \tag{4.17c}$$

where $\nu_1$ and $\tau_1$ are the barrier-shape and decay-rate correction factors for reference state R1, and $a^{\mathrm{nw}}$ is an universal constant given by

$$a^{\mathrm{nw}} \equiv z^{\mathrm{nw}} (e / g_{\mathrm{e}})^{3/2} \cong 1.079631 \times 10^{-10} \,\mathrm{A\,m^{-1}\,eV^{3/4}\,V^{-3/2}\,m^{3/2}} \tag{4.18}$$

Comparing (4.17b) with abstract form of the elementary FN-type equation (1.2) shows the following main differences: in the exponent, the zero-field barrier height $H_1$ is greater than the local thermodynamic work-function $\phi$ by the amount $W_1$, and a barrier-shape correction factor is present; in the pre-exponential, the power to which the decay width is raised is 3/2 rather than 2, and the term ($z^{\mathrm{nw}} w^{-1}$) has replaced $z_{\mathrm{S}}$. When, as in 4.17c, the $d_1^{3/2}$ term is expanded, $H_{\mathrm{R}}$ (rather than $\phi$) appears in the pre-exponential, the powers to which $F_{\mathrm{M}}$ and $H_{\mathrm{R}}$ are raised are different, a decay-rate correction factor appears and the "constant" terms are different.

Equation (4.17) is, of course, an elementary member of the family of nanowall CFE equations. Advanced members of the family will also include, in the pre-exponential, a tunnelling pre-factor and additional supply-type correction factors.

We emphasize that the differences between (4.17) and the elementary FN-type equation (1.2) result from differences in the way that the effective electron supply for the reference state has to be calculated when "lateral phase-locking" (and hence quantization of a lateral energy component) occur. With a thin nanowall, there are also special features associated with the shape of the tunnelling barrier. These are now explored.

### 5. Electrostatic potential variation for a classical nanowall

The expression for $J_L(F_M)$ contains parameters $G_n$ and $d_n$ that depend on the electrostatic potential $V(x,F_M)$ near the nanowall top surface. This potential could in principle be obtained by using a numerical Poisson solver, and $G_n$ and $d_n$ could in principle then be obtained by numerical integration of the JWKB integral. However, there are features of the mathematics of nanowalls that are not easily brought out by numerical solutions; thus, an analytical illustration is presented here. The first step obtains an expression for $V(x,F_M)$ by means of the conformal transformation illustrated in Fig. 3:

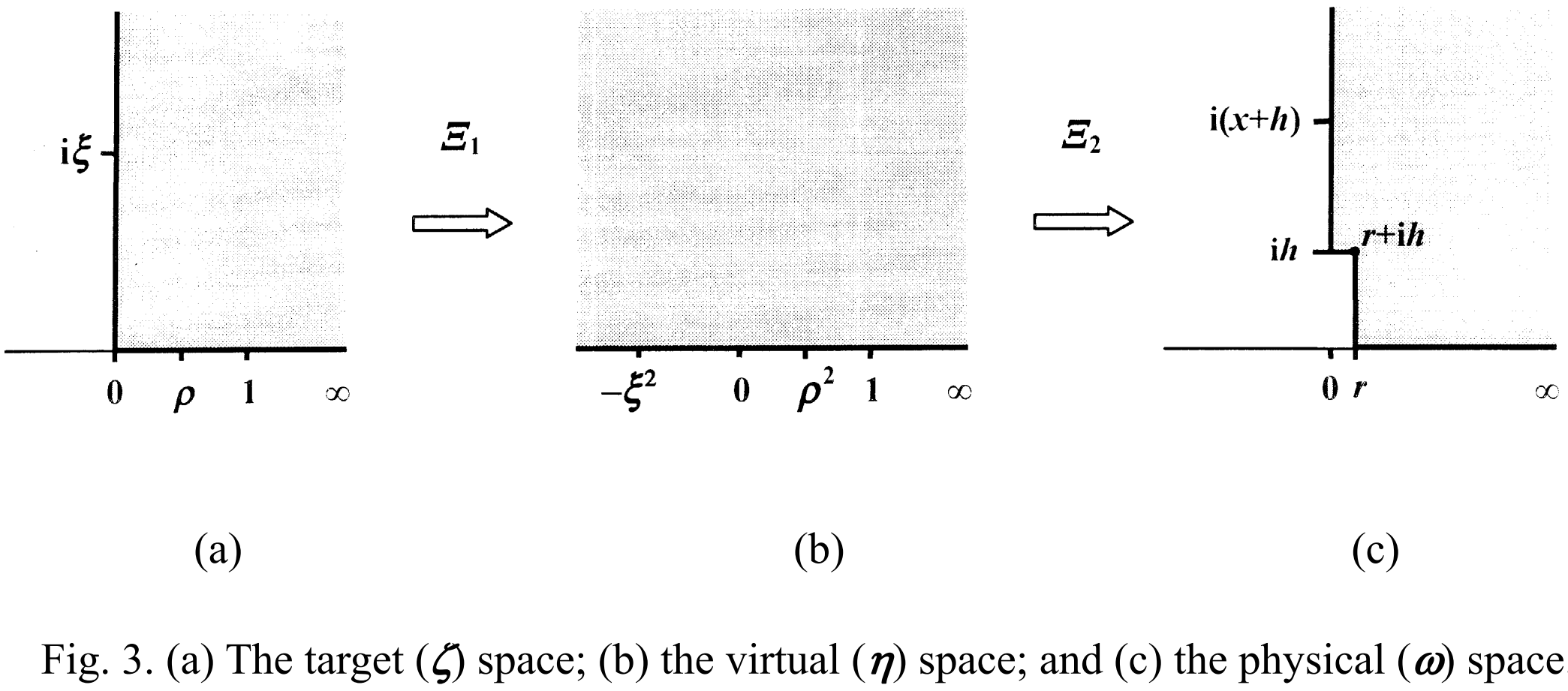

Fig. 3. (a) The target ($\zeta$) space; (b) the virtual ($\eta$) space; and (c) the physical ($\omega$) space. These spaces are connected via two conformal transformations.

The overall transformation $\Xi$ consists of two stages: $\Xi_1$ transforms the first quadrant in "target space" (Fig. 3a) into the positive half-plane in "virtual space" (Fig. 3b), and $\Xi_2$ then transforms this half-plane into a shape in "physical space" (Fig. 3c) that represents the right-hand half of the nanowall. Positions in the three spaces are represented by the complex numbers $\zeta$, $\eta$ and $\omega$, respectively, using axes as shown, with i≡√(−1).

In Fig 3c, $h$ is the height of the nanowall and $r$ is its half-width ($r$=$w$/2). The origin of $x$ is at 0+i$h$, and the position 0+i($h$+$x$) is an arbitrary point on the $x$-axis. The anode that applies the field is assumed to be located at an $x$-value ($x_a$=$X_a$–$h$) that is very large in comparison with $h$. The equivalences between the special points in the three spaces are shown in Table 2.

Table 2. Equivalent points in the three spaces in Fig. 3.

| | | | | |
|---|---|---|---|---|
| Target ($\zeta$) space | 0+i$\xi$ | 0 | $\rho$+i0 | 1+i0 |
| Virtual ($\eta$) space | $-\xi^2$+i0 | 0 | $\rho^2$+i0 | 1+i0 |
| Physical ($\omega$) space | i($x$+$h$) | 0+i$h$ | $r$+i$h$ | $r$+i0 |

To simplify equations, we use a mathematical function $E_S(\upsilon)$ defined in Appendix A($a$). Transformation $\Xi_1$ is simply: $\eta=\zeta^2$. Transformation $\Xi_2$ is the Schwarz–Christoffel transformation (Driscoll &Trefethen 2002).

$$\omega(\eta) = C\int_0^{\eta}\sqrt{\frac{\eta' - \rho^2}{\eta'(\eta' - 1)}}\,\mathrm{d}\eta' + \mathrm{i}h\,, \tag{5.1}$$

where $\eta'$ is a dummy variable. The values of $C$ in (5.1) and $\rho$ in Fig 3b are decided by the special points in Fig. 3c. These values are ultimately determined by the values of $r$ and the ratio $r/h$, via the relationships (see Appendix A($b$)):

$$\frac{r}{h} = \frac{\rho E_S(\rho)}{\sqrt{1-\rho^2}\, E_S(\sqrt{1-\rho^2})} \tag{5.2}$$

$$C = \frac{r}{2\rho E_S(\rho)} \tag{5.3}$$

From (5.2), one sees that $\rho$ is an implicit function of $(r/h)$. For practical field emitters, it is always true that $(r/h) \ll 1$. In this case, $\rho$ is a positive number given by

$$\rho \rightarrow 2\pi^{-1/2}(r/h)^{1/2}. \tag{5.4}$$

Here, and in equations below, the form of the relevant expression when $\rho$ is small is shown on the right of an arrow. For large values of $r/h$, $\rho$ becomes unity. For given $\rho$, a parameter $R$ is defined by

$$R = \rho E_S(\rho). \tag{5.5}$$

In the limit where $(r/h) \ll 1$ and hence $\rho \rightarrow 0$, $E_S(\rho) \rightarrow \pi\rho/4$ (see Appendix A(*a*)), and

$$R \rightarrow (r/h). \tag{5.6}$$

The point $0+\mathrm{i}\xi$ in target space corresponds to point $0+\mathrm{i}(x+h)$ in physical space. The value of $\xi$ is given implicitly in terms of $\rho$ and the ratio $x/r$ by (see Appendix A(*c*))

$$\xi\sqrt{\frac{1+\xi^2}{\rho^2+\xi^2}}-E(\theta\,|\,1-\rho^2)+\rho^2F(\theta\,|\,1-\rho^2)\;=\;Rx/r\,, \tag{5.7}$$

where $F(\varphi|m)$ and $E(\varphi|m)$ are the incomplete elliptic integrals of the first and second kinds, expressed in terms of amplitude $\varphi$ and elliptic parameter $m$ (see Appendix A($a$) for definitions), and $\theta$ is given by

$$\theta=\arcsin(\xi/\sqrt{\rho^2+\xi^2})\,. \tag{5.8}$$

In the target space, the anode is taken as represented by a straight line parallel to the real axis and at a distance $\xi_{\mathrm{a}}$ from it. In the physical space, the anode intersects the imaginary axis at the value $\mathrm{i}(h+x_{\mathrm{a}})$. The relationship between these two parameters is given by (see Appendix A($d$))

$$\frac{x_{\mathrm{a}}}{\xi_{\mathrm{a}}}\approx\lim_{\xi\to\infty}\frac{x}{\xi}=\frac{r}{R}\to h\,. \tag{5.9}$$

In the limit of large cathode-anode separation in the physical space, we have $x_{\mathrm{a}}>>h$ and $F_{\mathrm{M}}\approx V_{\mathrm{app}}/x_{\mathrm{a}}$ . Thus, the uniform "field" $F_\xi$ along the imaginary axis of the target space is the positive quantity:

$$F_\xi=\frac{V_{\mathrm{app}}}{\xi_{\mathrm{a}}}=\frac{V_{\mathrm{app}}}{x_{\mathrm{a}}}\frac{r}{R}\approx\frac{r}{R}F_{\mathrm{M}}\to hF_{\mathrm{M}}\,. \tag{5.10}$$

Note that, although $F_\xi$ has the function of a field, it formally has the dimensions of electrostatic potential.

It follows that in the target space the electrostatic potential at point 0+i$\xi$ is $F_\xi \xi$. Thus, in the physical space, the variation of electrostatic potential $V(x)$ along the $x$-axis is

$$V(x) = F_\xi \; \xi(x) \, , \tag{5.11}$$

where 0+i$\xi(x)$ is the point in the target space that corresponds to point 0+i($h$+$x$) in real space.

To proceed further, an expression for d$\xi$/d$x$ is needed. Appendix A($e$) shows that

$$\mathrm{d}\xi / \mathrm{d}x = (R/r)\sqrt{\frac{1+\xi^2}{\rho^2+\xi^2}} \; . \tag{5.12}$$

Using (5.9) and (5.10), the local field $F_{\mathrm{loc}}(x)$ in the physical space is given by

$$F_{\mathrm{loc}}(x) = \frac{\mathrm{d}V}{\mathrm{d}x} = \frac{\mathrm{d}V}{\mathrm{d}\xi}\frac{\mathrm{d}\xi}{\mathrm{d}x} = F_\zeta \frac{\mathrm{d}\xi}{\mathrm{d}x} = F_{\mathrm{M}}\sqrt{\frac{1+\xi^2}{\rho^2+\xi^2}} \; . \tag{5.13}$$

The Schwartz–Cristoffel transformation ensures that the point 0+i$h$ in real space maps into the point 0+i0 in target space. Thus, the local field ($F_{\mathrm{S}}$) at position S (i.e., at $x$=0) is obtained by putting $\xi$=0 into (5.13), and in the limit of small $\rho$ the field enhancement factor $\gamma_{\mathrm{S}}$ becomes

$$\gamma_{\mathrm{S}} = F_{\mathrm{loc}}(0) / F_{\mathrm{M}} = \frac{1}{\rho} \rightarrow \frac{\pi^{1/2}}{2}\left(\frac{h}{r}\right)^{1/2} \; . \tag{5.14}$$

The outer classical turning point at 0+i($h$+$x_n$) in real space corresponds to the point 0+i$\xi_n$ in target space, where $\xi_n$ is given by

$$\xi_n \equiv \xi(x_n) = H_n / eF_\xi = RH_n / eF_{\mathrm{M}} r \to H_n / eF_{\mathrm{M}} h. \tag{5.15}$$

To support this analytical treatment, solutions of Laplace's equation for the classical-nanowall geometry have been obtained by numerical (finite-element) methods (Qian, Wang and Li, unpublished work). Analytical and numerical solutions are in good agreement.

## 6. Detailed CFE theory for a classical nanowall

### *(a) The parameter $G_n$*

This section examines the parameters $G_n$ and $d_n$ that appear in expression (4.16) for the contribution $J_{\mathrm{L},n}(F_{\mathrm{M}})$ from sub-band SB$n$ to the total line current density. Starting from (4.7), and using (4.8), (5.11) and (5.12), we get

$$\begin{aligned} G_n(F_{\mathrm{M}}) &= g_{\mathrm{e}} \int_0^{x_n} \sqrt{H_n - eV(x, F_{\mathrm{M}})}\, \mathrm{d}x \;=\; g_{\mathrm{e}} H_n^{1/2} \int_0^{\xi_n} \frac{\sqrt{1-\xi/\xi_n}}{\mathrm{d}\xi/\mathrm{d}x} \mathrm{d}\xi \\ &= (r/R)(H_n)^{1/2} g_{\mathrm{e}} \int_0^{\xi_n} \sqrt{\frac{(1-\xi/\xi_n)(\rho^2+\xi^2)}{1+\xi^2}}\, \mathrm{d}\xi. \end{aligned} \tag{6.1}$$

Changing the variable of integration to $t = \xi / \xi_n$ yields

$$G_n(F_\mathrm{M}) = \{\xi_n r / R\}\ (H_n)^{1/2} g_\mathrm{e} \int_0^1 \sqrt{\frac{(1-t)(\rho^2 + \xi_n^2 t^2)}{1+\xi_n^2 t^2}}\, \mathrm{d}t\,. \tag{6.2}$$

We now introduce a parameter $\Gamma_n$ defined by

$$\Gamma_n^{-1} = \frac{3}{2}\int_0^1 \sqrt{\frac{(1-t)(\rho^2 + \xi_n^2 t^2)}{1+\xi_n^2 t^2}}\, \mathrm{d}t\,. \tag{6.3}$$

Equation (5.15) shows that the factor ($\xi_n r/R$) in (6.2) is equal to $H_n/eF_\mathrm{M}$. Hence

$$G_n(F_\mathrm{M}) \;=\; \frac{2g_\mathrm{e} H_n^{3/2}}{3e\Gamma_n F_\mathrm{M}} \;=\; bH_n^{3/2} / \Gamma_n F_\mathrm{M}\,, \tag{6.4}$$

where $b$ [$\equiv 2g_\mathrm{e}/3e$] is the second FN constant, as before.

*(b) The parameter $d_n$*

From (4.9), using substitutions similar to those in Section 6.1:

$$\begin{aligned} d_n^{-1} &= (g_\mathrm{e}/2)\int_0^{\xi_n} [H_n - eV(x)]^{-1/2}\, \mathrm{d}x \\ &= \frac{r}{R} H_n^{-1/2} g_\mathrm{e} \int_0^{\xi_n} \sqrt{\frac{\rho^2+\xi^2}{(1-\xi/\xi_n)(1+\xi^2)}}\, \mathrm{d}\xi. \end{aligned} \tag{6.5}$$

Changing the variable of integration to $t = \xi/\xi_n$ yields

$$d_n^{-1} = \{\xi_n r / R\} H_n^{-1/2} g_e \int_0^1 \sqrt{\frac{\rho^2 + \xi_n^2 t^2}{(1-t)(1+\xi_n^2 t^2)}} \, \mathrm{d}t \,. \tag{6.6}$$

A parameter $\Theta_n$ may be introduced by

$$\Theta_n^{-1} \equiv \frac{1}{2} \int_0^1 \sqrt{\frac{\rho^2 + \xi_n^2 t^2}{(1-t)(1+\xi_n^2 t^2)}} \, \mathrm{d}t \,. \tag{6.7}$$

Putting $(\xi_n r/R) = (H_n/eF_M)$, as above, yields

$$d_n = \Theta_n (e / g_e) H_n^{-1/2} F_M \,. \tag{6.8}$$

*(c) Limiting expressions for $\Gamma_n$ and $\Theta_n$*

In general, the various correction factors $\Gamma_n$ and $\Theta_n$ need to be calculated numerically. However, useful approximate expressions exist because the term $(\rho^2 + \xi_n^2 t^2)^{1/2}$ that appears in the integrands in (6.3) and (6.7) can take different limiting forms, depending on the relative sizes of $\rho$ and $\xi_n t$.

*(c)(i) The "slowly varying field" approximation: $\xi_n << \rho$*

In the limit where $\xi_n << \rho$, (which corresponds physically to the case where the barrier length $x_n$ is small in comparison with nanowall half-width $r$), it can be shown by using a computer algebra

package that

$$\mathit{\Gamma}_n^{-1} \approx \frac{3\rho}{2}\int_0^1 \sqrt{\frac{(1-t)}{1+\xi_n^2 t^2}}\,\mathrm{d}t \approx \rho \approx \gamma_\mathrm{S}^{-1}\,, \tag{6.9}$$

$$\mathit{\Theta}_n^{-1} \approx \frac{1}{2}\rho\int_0^1 \sqrt{\frac{t^2}{(1-t)(1+\xi_n^2 t^2)}}\,\mathrm{d}t \approx \rho \approx \gamma_\mathrm{S}^{-1} \tag{6.10}$$

We call this the "slowly varying field" approximation. In this limit, the decay of the field strength with distance along the $x$-axis is relatively slow, and the barrier is "nearly triangular". Thus, the barrier-shape correction factor $\nu_n \approx 1$, the decay-rate correction factor $\tau_n \approx 1$, and the correction factors $\mathit{\Gamma}_n$ and $\mathit{\Theta}_n$ both become effectively equal to the field enhancement factor $\gamma_\mathrm{S}$.

*(c)(ii) The "sharply varying field" approximation: $\rho << \xi_n$*

The opposite limit $\rho << \xi_n$ corresponds physically to the situation where the nanowall half-width $r$ is small in comparison with the barrier length $x_n$. We call this the "sharply varying field" approximation. In this limit, it can be shown by using a computer algebra package and eq. (5.15) that

$$\mathit{\Gamma}_n^{-1} \approx \frac{3}{2}\xi_n\int_0^1 \sqrt{\frac{(1-t)t^2}{1+\xi_n^2 t^2}}\,\mathrm{d}t \approx \frac{2\xi_n}{5} \approx \frac{2H_n}{5eF_\mathrm{M}h}\,, \tag{6.11}$$

$$\Theta_n^{-1} \approx \frac{\xi_n}{2}\int_0^1 \frac{t}{\sqrt{(1-t)(1+\xi_n^2 t^2)}}\mathrm{d}t \approx \frac{2\xi_n}{3} \approx \frac{2H_n}{3eF_\mathrm{M}h}. \tag{6.12}$$

Note that the issue of which of these approximations might be more appropriate in a given situation depends both on the nanowall half-width $r$, and on the strength of the field near the surface, since the latter determines the width $x_n$ of the tunnelling barrier.

*(d) CFE equations*

If the nanowall is sufficiently thin (i.e., if $w$ is sufficiently small that $(G_2–G_1)>1$), then emission from the $n$=1 sub-band is dominant, and (by using (4.17)) the characteristic area current density $J_\mathrm{A}(F_\mathrm{M})$ for the nanowall can be approximated by

$$J_\mathrm{A}(F_\mathrm{M}) = J_\mathrm{L}(F_\mathrm{M})/w \approx a^\mathrm{nw} w^{-1}\Theta_1^{3/2} H_1^{-3/4} F_\mathrm{M}^{3/2} \exp[-bH_1^{3/2} / \Gamma_1 F_\mathrm{M}] \tag{6.13}$$

Comparisons show that $\Gamma_1$ is a generalised replacement for the combination $\gamma_\mathrm{S}/\nu_1$ in (4.17) , and $\Theta_1$ a generalised repacment for $\gamma_\mathrm{S}/\tau_1$.

In the "slowly varying field" limit, (6.9) and (6.10) show that a CFE equation is obtained from (6.13) by replacing both $\Gamma_1$ and $\Theta_1$ by $\gamma_\mathrm{S}$:

$$J_\mathrm{A}(F_\mathrm{M}) \approx a^\mathrm{nw} w^{-1} H_1^{-3/4} \gamma_\mathrm{S}^{3/2} F_\mathrm{M}^{3/2} \exp[-bH_1^{3/2} / \gamma_\mathrm{S} F_\mathrm{M}]. \tag{6.14}$$

This is simply (4.17) with $\nu_1$=1, $\tau_1$=1; $\gamma_S$ is given by (5.14).

In the "sharply varying field" limit, (6.11) and (6.12), with $n$=1, can be used to substitute for $\Gamma_1$ and $\Theta_1$ in (6.13), leading to

,

$$J_A(F_M) \approx a^{svf}(h^{3/2}/w)H_1^{-9/4}F_M^3\exp[-\tfrac{2}{5}bH_1^{5/2}/hF_M^2], \qquad (6.15)$$

where $a^{svf}$ is an universal constant given by

$$a^{svf} \cong 1.983410\times10^{-10}\ \mathrm{A\ m^{-5/2}\ eV^{9/4}\ V^{-3}\ m^3}. \qquad (6.16)$$

Equation (6.15) again has the form of (4.17), with $\gamma_S$ given by (5.14), but here $\nu_1$ and $\tau_1$ are functions of $H_1$ and $F_M$ (or $F_S$), given by

$$\nu_1 \approx (\pi^{1/2}/5)H_1/eF_M r^{1/2}h^{1/2} = (\pi/10)H_1/eF_S r, \qquad (6.17)$$

$$\tau_1 \approx (\pi^{1/2}/3)H_1/eF_M r^{1/2}h^{1/2} = (\pi/6)H_1/eF_S r. \qquad (6.18)$$

Clearly, $\nu_1$ and $\tau_1$ tend to become large as $r$ becomes small for constant $h$ and $F_M$, or as $F_S$ becomes small for constant $r$, or as $F_M$ becomes small for constant $r$ and $h$. In the opposite limit, formulae (6.17) and (6.18) break down as we move back towards the "slowly-varying field" limit, and $\nu_1$ and $\tau_1$ approach unity.

### *(e) Comment*

Obviously, (6.14) would generate a nearly-straight line in FN plot or a Millikan-Lauritsen (ML) plot, but the power of $F_M$ in the exponent is slightly different from that predicted by the elementary FN-type equation. This difference underlines the merit (when the FN plot or ML plot is nearly straight) of attempting to measure the empirical power of $F_M$ (or, equivalently, applied voltage) in the pre-exponential, as proposed elsewhere (Forbes 2008a).

Equation (6.15), on the other hand, would generate a curved line in a FN or ML plot, but would generate a straight line in a plot of type [$\ln\{J_A\}$ vs $1/V_{app}^2$].

## 7. Discussion

### *(a) Summary*

This paper has stated (in Section 2) a general methodology for developing theoretical CFE equations, and has applied it to CFE from a thin nanowall. A feature of this work is the complexity of the emission phenomena uncovered, despite the use of many simplifying assumptions (equivalent to those used by FN and others when analyzing CFE from bulk metals).

A major feature results from the "quantum confinement" in the direction of the nanowall width (i.e., phase-locking of the relevant wave-function component, and consequent quantization of the associated energy component $W_z$). This requires the electron states in the nanowall to be grouped into sub-bands defined by a quantum number $n$ that determines the allowed values of $W_z$. Also, due to the need to "use up" some of its total energy to provide this lateral quantization energy, an electron at the Fermi level sees a tunnelling barrier of zero-field height greater than the local thermodynamic work-function.

Options within the theory depend on the absolute size of the nanowall half-width $r$, and on the relative sizes of the nanowall height $h$, its half-width $r$, and the lengths $x_n$ of the tunnelling barriers for the references states (R$n$) associated with the various sub-bands. Also, values of $x_n$ are influenced by the value of the applied macroscopic field $F_M$ or voltage $V_{app}$.

In respect of electron supply, if $r$ is sufficiently small, then the theory can be simplified by assuming that only the lowest sub-band contributes significantly to electron emission. This has been designated the "thin nanowall" case. The physical condition assumed here is that the tunnelling probability for emission from the reference state R2 should be less that than for reference state R1 by a factor of at least e (≈2.7), which requires $(G_2–G_1)>1$. Emission from the $n$=1 sub-band would also be dominant if the states in higher sub-bands were unoccupied, but we consider the tunnelling-probability condition to be more critical.

A second effect is produced by the possible influence of a sharp emitter on the shape of the tunnelling barrier, in particular on its influence on how quickly the local electric field decays with distance from the emitting surface, over a distance comparable with the length of the tunnelling barrier. In the thin nanowall case, if conditions are such that this decay is slow (so the barrier is nearly exactly triangular), then (6.14) applies; if conditions are such that this decay is fast (as illustrated in Fig. 2), then (6.15) applies.

*(b) Conditions of applicability*

Because of (a) interconnections between different parts of the theory, (b) the absence of clear transitions and (c) the existence of other constraints, it is difficult to formulate reliable, precise algebraic or numerical conditions for the clear existence of the emission situations discussed above. The following treatment aims to be indicative rather than exact.

The issue of dominance of emission from the $n$=1 sub-band depends mainly on the nanowall width. The condition $(G_2–G_1)>1$ requires:

$$(b / F_{\mathrm{M}})(H_2^{3/2} / \Gamma_2 - H_1^{3/2} / \Gamma_1) > 1. \tag{7.1}$$

If, in addition, conditions are such that the "slowly decaying field approximation" applies, then $\Gamma_2=\Gamma_1=\gamma_{\mathrm{S}}$, and (7.1) becomes $(b / F_{\mathrm{S}})(H_2^{3/2} - H_1^{3/2}) > 1$. Using (4.8) to write $H_n=\phi+n^2W_{z1}$, expanding this by the binomial theorem, and using (3.3) for $W_{z1}$, we get the requirement

$$r^2 < \frac{9}{64}\frac{h_{\mathrm{P}}^2}{m_{\mathrm{e}}}\frac{b\phi^{1/2}}{F_{\mathrm{S}}} \approx (2.889698 \text{ eV}^{-1/2} \text{ V nm}^{-1}) \times \frac{\phi^{1/2}}{F_{\mathrm{S}}} \tag{7.2}$$

$\phi$ can be taken as 5 eV. It is difficult to guess $F_{\mathrm{S}}$ accurately, so we assume $F_{\mathrm{S}}$~5 V/nm. This yields the condition $r$<1.1 nm.

If emission from the $n$=1 sub-band is dominant, then the condition for the "slowly varying field" approximation to be the better approximation is $\rho>\xi_1$. Using (1.1), (4.16) and (5.15), we obtain

$$r > \frac{\pi}{4}\frac{H_1}{eF_S}. \tag{7.3}$$

Approximating $H_1 \sim \phi \approx 5$ eV, $F_S \sim 5$ V/nm, yields $r > 0.8$ nm.

These results suggest that (for given values of $h$ and $F_S$) there is a narrow range of $r$-values near 1 nm where emission from the $n$=1 sub-band is dominant, and where we expect a FN or ML plot to be nearly straight (but where we expect the zero-field barrier height $H_1$ to be greater than the local thermodynamic work-function $\phi$). For larger $r$-values, there is a slow transition to Fowler-Nordheim-like behaviour. For smaller $r$-values, there is a transition towards behaviour described by (6.15). However, one might expect the use of a square potential well (to describe the lateral quantum-confinement behaviour) to become an inadequate model when $r$ approaches atomic dimensions.

In principle, in such circumstances, one should develop atomic-theory-based quantum-mechanical models for the band-structure of a thin slab comprising several layers of atoms, but this is beyond the scope of the present paper. When the width of the nanowall is a few atoms, then one might expect (6.15) to be unreliable in detail. However, the effects associated with the barrier shape relate to the electrostatics of the field distribution above the nanowall; thus, they should physically occur for nanowalls with half-width of order 1 nm or less, whatever quantum-mechanical model one uses for the thin nanowall. Hence, there is a more general expectation that, for nanowall emitters of this size, FN and ML plots may be curved.

*(c) Conclusions*

The main conclusions from this work are as follows. In the CFE regime, the behaviour of a nanowall field emitter of sufficiently small width will depart from the behaviour described by FN-type equations. The present work suggests that a suitable criterion for significant departure from FN-type behaviour might be a width $w$ of around 2 nm, or below.

When an emitter is as thin as this, there are two intrinsically linked possible consequences: quantum confinement effects, and barrier effects that occur because the local field falls off rapidly with distance from the surface (hence the electrostatic PE variation becomes "cusp-like"). Detailed consequences of having quantum confinement and a cusp-like barrier, in particular the consequences for the form of theoretical CFE equations that give emission current density, have been described.

The best experimental approach for detecting the effects described here may be to look, first for curvature in measured FN or ML plots (which is a known experimental phenomenon with carbon field emitters, but which might have other causes), and then for current vs voltage behaviour that specifically conforms with (6.15).

Although there is previous work that investigates either cusp-like barrier effects (e.g., Cutler et al. 1993, Edgcombe & de Jonge 2006) or quantum-confinement effects for small-sized emitters (for example, Chen and Liang 2008), we believe that this paper is the first work that looks at the operation of both types of effect together, for a particular emitter geometry (the thin nanowall). We have felt that the conceptual structure of the physical results would be best brought out by using a relatively naive analytical model. Clearly there is scope both for detailed numerical investigations that extend and amplify the present work, and for related investigations based on detailed atomic-level models of the band-structure.

More generally, we consider that it will be important to co-investigate quantum confinement and barrier-shape effects for other experimentally plausible emitter geometries. Expectation is that emission from such emitters may conform to emission equations that are both different from FN-type equations, and different in detail from the equations presented here. What we expect, eventually, is that CFE will be described by several families of theoretical equations, with each family corresponding to a particular type of quantum-confinement arrangements (nanowall, post, thin-slab, localized-state, etc.). Fowler-Nordhim-type equations will be the family that applies when no significant quantum-confinement effects occur.

The present work also underlines the need to make a clear distinction in CFE theory between the concepts of (a) local thermodynamic work-function $\phi$ and (b) zero-field barrier height $H_R$ for a particular reference state.

ACKNOWLEDGMENTS

The project is supported by the National Natural Science Foundation of China (Grant Nos. 10674182, 10874249, and 90306016) and the National Basic Research Programme of China (2007CB935500 and 2008AA03A314).

## Appendix A Mathematical details

*(a) Elliptic integrals and related functions*

The incomplete elliptic integrals of the first and second kinds, $F(\varphi|m)$ and $E(\varphi|m)$ respectively, are defined in terms of the elliptic parameter $m$ and the amplitude $\varphi$ by

$$F(\varphi\,|\,m) = \int_0^{\varphi} (1 - m\sin^2\vartheta)^{-1/2}\,\mathrm{d}\vartheta\,, \tag{A.1}$$

$$E(\varphi\,|\,m) = \int_0^{\varphi} (1 - m\sin^2\vartheta)^{+1/2}\,\mathrm{d}\vartheta\,. \tag{A.2}$$

The complete elliptic integrals of the first and second kinds, $K(m)$ and $E(m)$ respectively, are defined by: $K(m) = F(\tfrac{\pi}{2}\,|\,m)$; $E(m) = E(\tfrac{\pi}{2}\,|\,m)$.

It is convenient, here, to introduce a special function $E_S(\upsilon)$, related to $E(\varphi|m)$ and defined by

$$\begin{aligned} E_{\mathrm{S}}(\upsilon) &= E(\arcsin(\upsilon)\,|\,\upsilon^{-2}) \\ &= \int_0^{\arcsin(\upsilon)} (1 - \upsilon^{-2}\sin^2\vartheta)^{1/2}\,\mathrm{d}\vartheta. \end{aligned} \tag{A.3}$$

When $\upsilon \to 0$, then $\arcsin(\upsilon) \to \upsilon$ and the range of integration in $\vartheta$ becomes small. Then:

$$E_{\mathrm{S}}(\upsilon) \;\approx\; \int_0^{\upsilon} \sqrt{1 - \frac{\vartheta^2}{\upsilon^2}}\,\mathrm{d}\vartheta \;=\; \frac{\pi}{4}\upsilon\,. \tag{A.4}$$

*(b) Schwatz–Christoffel transformations*

Here, the integral in the $\Xi_2$ transformation (eq. (43)) is calculated. It is a contour-integral in the virtual ($\eta$) space, as illustrated in Fig. 4. There are two singular points, at the origin and at $\eta$ =1, respectively. Two infinitely small half-circles, around the origin (C0) and around $\eta$ =1 (C1), have been introduced in order to give the integral definite meaning. The infinitely small positive number $\varepsilon$ is set to zero after calculations have been carried out.

The integral eq. (43) is thereby understood as a principal integral:

$$\omega(\eta) = C \int_{\substack{\varepsilon \\ (\Sigma)}}^{\eta} \sqrt{\frac{\eta' - \rho^2}{\eta'(\eta' - 1)}} \, \mathrm{d}\eta' + ih \quad , \tag{A.5}$$

along the contour Σ, then putting $\varepsilon \to 0$. It can be verified that the integrals around the two half-circles C0 and C1 become zero in this limit.

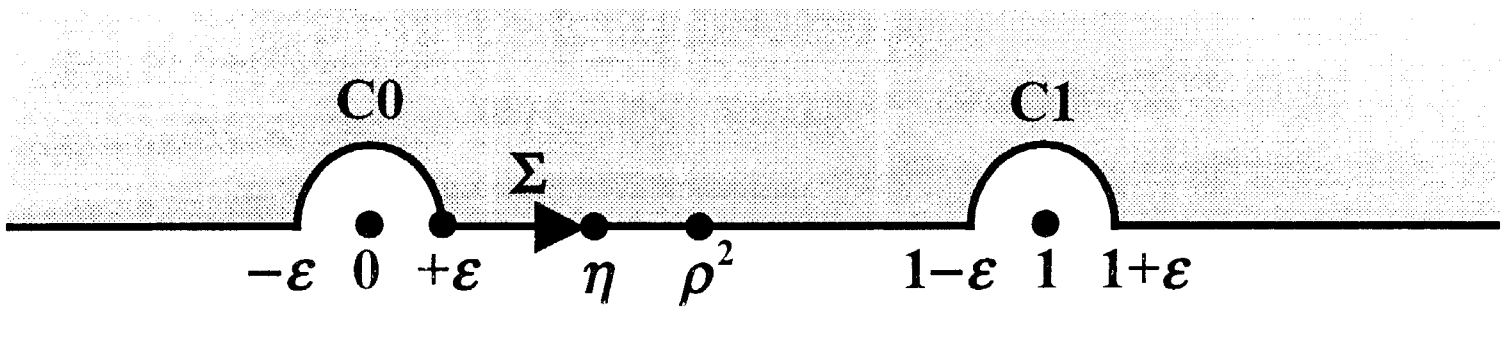

Fig. 4. The contour Σ for the transformation $\Xi_2$ lies along the real axis in $\eta$-space.
In effect, it starts at the position $+\varepsilon$ and finishes at the position $\eta$.

According to Table 2, for $\omega = r + \mathrm{i}h$ the transformation (A.5) becomes

$$r = C \int_{\substack{\varepsilon \\ (\Sigma)}}^{\rho^2} \sqrt{\frac{\eta' - \rho^2}{\eta'(\eta' - 1)}} \, \mathrm{d}\eta' \, . \tag{A.6}$$

Let $C$ be a real number. Since the contour $\Sigma$ lies along the real axis, the variable $\eta'$ in (A.6) is real and positive, and lies in the range $\varepsilon \le \eta' \le 1$. Since $r$ is also real, $\eta' - \rho^2$ and $\eta' - 1$ must have the same sign. Therefore $\rho < 1$, and (A.6) can be written as an ordinary integral

$$r = C \int_0^{\rho^2} \sqrt{\frac{\rho^2 - t}{t(1-t)}} \mathrm{d}t \,, \tag{A.7}$$

where $\eta'$ has been replaced by $t$. On replacing $t$ by $\sin^2 \vartheta$, and using (A.3), one has

$$r = 2\rho C \int_0^{\arcsin \rho} \sqrt{1 - \rho^{-2} \sin^2 \vartheta} \mathrm{d}\vartheta = 2\rho C E_{\mathrm{S}}(\rho) \,. \tag{A.8}$$

From this, equation (5.3) for the constant $C$ is established. On defining $R \equiv \rho E_{\mathrm{S}}(\rho)$, we obtain

$$C = \frac{r}{2R} \,. \tag{A.9}$$

At $\omega = r + \mathrm{i}0$, one has

$$\begin{aligned} r &= C \int_{\substack{\varepsilon \\ (\Sigma)}}^{1} \sqrt{\frac{\eta' - \rho^2}{\eta'(\eta' - 1)}} \mathrm{d}\eta' + \mathrm{i}h \\ &= C \int_{\substack{\varepsilon \\ (\Sigma)}}^{\rho^2} \sqrt{\frac{\eta' - \rho^2}{\eta'(\eta' - 1)}} \mathrm{d}\eta' + C \int_{\substack{\rho^2 \\ (\Sigma)}}^{1} \sqrt{\frac{\eta' - \rho^2}{\eta'(\eta' - 1)}} \mathrm{d}\eta' + \mathrm{i}h \end{aligned} \tag{A.10}$$

On using (A.6) and recalling that $\rho < 1$, this can be written as an equation involving the ordinary integral

$$-\mathrm{i}C \int_{\rho^2}^{1} \sqrt{\frac{t' - \rho^2}{t'(1 - t')}} \mathrm{d}t' + \mathrm{i}h = 0 \,. \tag{A.11}$$

On substituting $t' = 1 - t$, this becomes

$$h = C \int_0^{1-\rho^2} \sqrt{\frac{1 - \rho^2 - t}{t(1-t)}} \mathrm{d}t \,. \tag{A.12}$$

Integral (A.12) has the same form as (A.7), except that $\rho^2$ is replaced by $1-\rho^2$. Hence,

$$h = 2C\sqrt{1 - \rho^2}\, E_{\mathrm{S}}(\sqrt{1 - \rho^2}) \,. \tag{A.13}$$

Thus, using (A.9), one obtains

$$\frac{r}{h} = \frac{R}{\sqrt{1-\rho^2}\, E_S(\sqrt{1-\rho^2})} . \tag{A.14}$$

When $R$ is replaced by its definition (5.5), (A.14) becomes equation (5.2).

*(c) Relation involving the ratio x/r*

At $\omega$=0+i($x$+$h$), (A.5) gives

$$\mathrm{i}x = C \int_{\substack{-\varepsilon \\ (\Sigma)}}^{-\xi^2} \sqrt{\frac{\eta' - \rho^2}{\eta'(\eta' - 1)}}\, \mathrm{d}\eta' = -\frac{C}{\mathrm{i}} \int_0^{\xi^2} \sqrt{\frac{t+\rho^2}{t(t+1)}}\, \mathrm{d}t . \tag{A.15}$$

In the last step, we introduced the dummy variable $t = -\eta'$. An integral $I$ can then be defined by

$$I \equiv \int_0^{\xi^2} \sqrt{\frac{t+\rho^2}{t(t+1)}}\, \mathrm{d}t = \frac{x}{C} . \tag{A.16}$$

On changing the dummy variable to $s = \sqrt{t/(t+\rho^2)}$, and defining $m = 1 - \rho^2$ and $s_0 = \xi / \sqrt{\rho^2 + \xi^2}$, we obtain

$$I = \int_0^{s_0} \frac{2\rho^2 \mathrm{d}s}{(1-s^2)\sqrt{(1-s^2)(1-ms^2)}} . \tag{A.17}$$

Note that $0 < s_0 < 1$ and $0 < m < 1$. Because

$$\begin{aligned} &\frac{1}{(1-s^2)\sqrt{(1-s^2)(1-ms^2)}} \\ &= \frac{\mathrm{d}}{\mathrm{d}s}\left[\frac{s}{1-m}\sqrt{\frac{1-ms^2}{1-s^2}}\right] - \frac{1}{1-m}\sqrt{\frac{1-ms^2}{1-s^2}} + \frac{1}{\sqrt{(1-s^2)(1-ms^2)}} \end{aligned} , \tag{A.18}$$

(A.17) can be written as

$$I = 2\rho^2 \left[ \frac{s_0}{1-m}\sqrt{\frac{1-ms_0^2}{1-s_0^2}} - \frac{1}{1-m}\int_0^{s_0}\sqrt{\frac{1-ms^2}{1-s^2}}\,\mathrm{d}s + \int_0^{s_0}\frac{\mathrm{d}s}{\sqrt{(1-s^2)(1-ms^2)}} \right]. \tag{A.19}$$

On putting $s = \sin\vartheta$ and $\theta = \arcsin(s_0)$, (A.19) becomes

$$I = 2\rho^2 \left[ \frac{s_0}{1-m}\sqrt{\frac{1-ms_0^2}{1-s_0^2}} - \frac{1}{1-m}\int_0^{\theta}\sqrt{1-m\sin^2\vartheta}\,\mathrm{d}\vartheta + \int_0^{\theta}\frac{\mathrm{d}\vartheta}{\sqrt{1-m\sin^2\vartheta}} \right]. \tag{A.20}$$

On using definitions (A.1) and (A.2), we obtain

$$I = 2\rho^2 \left[ \frac{s_0}{1-m}\sqrt{\frac{1-ms_0^2}{1-s_0^2}} - \frac{1}{1-m}E(\theta\,|\,m) + F(\theta\,|\,m) \right]. \tag{A.21}$$

On substituting eq. (A.21) into (A.16), and using the definitions for $m$ and $s_0$, we obtain

$$\xi\sqrt{\frac{1+\xi^2}{\rho^2+\xi^2}} - E(\theta\,|\,1-\rho^2) + \rho^2 F(\theta\,|\,1-\rho^2) = \frac{x}{2C}. \tag{A.22}$$

The definition for $\theta$ is the same as (5.8). On replacing $C$ by using (A.9), we obtain (5.7), which is an implicit relationship between $\xi$, $\rho$ and $x/r$.

### *(d) Relationship between $x_a$ and $\xi_a$*

For $\rho<1$, in the limit $\xi \to \infty$, both $E(\theta\,|\,1-\rho^2)$ and $\rho^2 F(\theta\,|\,1-\rho^2)$ are finite. From (A.22), one has

$$\lim_{\xi\to\infty}\sqrt{\frac{1+\xi^2}{\rho^2+\xi^2}} = \lim_{\xi\to\infty}\frac{x}{2C\xi}. \tag{A.23}$$

The left hand side is 1. Hence

$$\lim_{\xi\to\infty}\frac{x}{\xi} = 2C = \frac{r}{R}. \tag{A.24}$$

where (A.9) has been used.

*(e) Expression for* d$\xi$/d$x$

For given values of $\rho$ and $C$, integral equation (A.16) implies the differential equation

$$\sqrt{\frac{\xi^2+\rho^2}{\xi^2(\xi^2+1)}}2\xi \mathrm{d}\xi = \frac{\mathrm{d}x}{C} . \qquad \text{(A.25)}$$

On substituting result (A.9) into (A.25), we obtain (for positive $\xi$) that

$$\mathrm{d}\xi / \mathrm{d}x = (R/r)\sqrt{\frac{1+\xi^2}{\rho^2+\xi^2}} . \qquad \text{(A.26)}$$

This is equation (5.12).